\documentstyle[psfig,conf_iap,10pt]{article}
\begin{document}

\heading{Reionization Feedback and the Photoevaporation of Intergalactic
Clouds} 

\par\medskip\noindent

\author{Paul R. Shapiro,$^{1,2}$ Alejandro C. Raga,$^2$ and
Garrelt Mellema$^3$}

\address{Department of Astronomy, University of Texas, Austin, TX 78712, USA}

\address{Instituto de Astronom\'\i a-UNAM, Apdo Postal 70-264,
04510 M\'exico D. F., M\'exico}

\address{Stockholm Observatory, S-133 36 SaltsJ\"obaden, Sweden}

\begin{abstract}
Energy released by a small fraction of the baryons in the universe, 
which condensed out of while the IGM was cold, dark, and neutral,
reheated and reionized it, exposing gas clouds within it
to the glare of ionizing radiation. The first gas dynamical simulations of
the photoevaporation of an intergalactic cloud by a quasar, including radiative
transfer, are presented, along with a few observational diagnostics.
\end{abstract}

\section{Reionization, Reheating, and Feedback}
The Gunn-Peterson limit implies that the IGM was reionized and reheated
by $z\sim5$, by energy released by objects which previously
condensed out of the background and formed stars, AGNs or some other sources.
This exerted a negative feedback on the rate of collapse of gas out of the
background IGM by raising the Jeans mass there and affected the appearance and
evolution of the IGM and the structure which subsequently collapsed out of it.
How early could starlight have reionized the IGM? Shapiro and Giroux 
attempted to answer this question by solving linear equations for 
the growth of density fluctuations in the IGM and 
using a Press-Schechter approach
to determine the baryon fraction which had collapsed out at
any epoch, releasing energy to reheat and reionize the IGM, coupled
to a detailed numerical evolution of the thermal and ionization balance
of a coarse-grained, spatially-averaged IGM and the equation of radiative
transfer for the ionizing radiation background,
including the opacity of the observed quasar absorption line gas,
as described in \cite{X}, \cite{X2}, and \cite{X3}.
The maximum-possible efficiency was assumed
for energy release by massive stars, which form at a rate proportional
to the rate of collapse of baryons out of the IGM and stop forming when they 
have enriched 
the collapsed fraction with a solar abundance of heavy elements,
in a flat, COBE-normalized, standard CDM model ($\Omega_B=0.06$, 
$h=0.5$, $\Omega=1$). The IGM was found to reheat by $z_h\sim65$ to
$T_{\rm IGM}\sim10^{3.5}{\rm K}$ when the collapsed baryon fraction was
only $f_{\rm coll}\approx10^{-4}$, while the H atom ionization
breakthrough was $z_b\sim50$ when $f_{\rm coll}\sim10^{-3}$, implying
a net metallicity averaged over all baryons in the universe at these epochs 
equal to these values of $f_{\rm coll}$, in solar units
\cite{X3}. Since the COBE-normalization
is almost twice that typically assumed in recent simulations of the Lyman 
alpha forest and galaxy formation, our results are {\it conservative}
in the sense of maximizing $z_b$;
a lower initial amplitude would make reionization  and reheating occur
a little later than this. In my talk, I also presented
ASPH simulations by Shapiro and Martel 
which demonstrated that
global reheating can significantly affect
the small-scale structure formation responsible
for quasar absorption line gas(\cite{X3},\cite{X4},\cite{X5}). 
This contribution is too
brief to present that material, but I refer the reader to \cite{X3} for
a summary. Here we will focus, instead, on new work on
the effects of photoionization.

\section{The Photoevaporation of Intergalactic Clouds}

The first sources of ionizing radiation which turned on in the neutral 
(i.e. postrecombination) IGM prior to $z\sim5$ resulted in isolated,
expanding H~II regions. The expansion and eventual overlap
of the weak, R-type cosmological 
ionization fronts bounding these H~II regions was 
previously described
analytically by treating the IGM as a uniform, cosmologically expanding gas
(\cite{X6},\cite{X7}). The density fluctuations required to explain
galaxy formation and the Lyman alpha forest were accounted for
approximately by treating the IGM as ``clumpy,'' with a universal clumping
factor. That approximation is correct in the limit in which the clumps are
either too small to ``self-shield'' or else constitute only a small
fraction of the total mass inside the HII region. It does not, however,
address the possible dynamical consequences for the clumps, themselves, of
the passage of I-fronts. In what follows, we present the first simulations
of the gas dynamics and radiative transfer of an intergalactic cloud
overtaken by a cosmological I-front.

The fate of this cloud depends  fundamentally on whether or not it can shield
itself against the incident radiation from the external source responsible
for the intergalactic I-front. If the cloud size exceeds the 
``Str\"omgren length'' (the length of a column of gas 
within which the unshielded arrival rate of ionizing photons just balances the 
total recombination rate), it can trap the I-front. In that case,
the weak R-type I-front which swept into the cloud initially, 
moving supersonically with respect to
gas both ahead and behind, decelerates to the sound speed of
the ionized gas before it can exit the cloud, thereby
becoming a weak, D-type front preceded by a shock.
Typically, the side of the cloud which faces the radiation source expels
a supersonic wind which causes the remaining cloud material to be accelerated
away from the source by the so-called ``rocket effect'' as the cloud
photoevaporates (cf.\cite{X8}). For a uniform gas of H 
density $n_{\rm H,c}$, located 
a distance $r_{\rm Mpc}$ (in Mpc) from a UV source
emitting $N_{\rm ph,56}$ ionizing photons per second (in units of 
$10^{56}{\rm s}^{-1}$), the Str\"omgren length is only 
$\ell_{\rm S}\cong(50\,{\rm pc})(N_{\rm ph,56}/r_{\rm Mpc}^2)
(n_{\rm H,c}/0.1\,{\rm cm}^{-3})^{-2}$. Gas bound
to dark matter halos whose virial temperature is less than 
$10\,{\rm km\,s^{-1}}$ will photoevaporate unimpeded by gravity. 
For larger halos gravity competes with the effects of photoevaporation.

As a first study of these important effects, we have simulated the
photoevaporation of a uniform, spherical, neutral, intergalactic cloud of 
gas mass $1.5\times10^6M_\odot$, radius $R_c=0.5\,\rm kpc$, 
density $n_{\rm H,c}=0.1\,{\rm cm^{-3}}$ and $T=100\,\rm K$,
in which self-gravity is unimportant, located 
$1\,\rm Mpc$ from a quasar with emission spectrum $F_\nu\propto\nu^{-1.8}$
($\nu>\nu_{\rm H}$) and $N_{\rm ph}=10^{56}{\rm s}^{-1}$, initially
in pressure balance with
an ambient IGM of density $0.001\,\rm cm^{-3}$ which
at time $t=0$ has
just been photoionized by the passage of
the intergalactic R-type I-front generated when the quasar turned on.
Apart from H and He, the cloud also contains heavy elements at $10^{-3}$
times the solar abundance. Our simulations in 2D, axisymmetry use an
Eulerian hydro code (called CORAL), with Adaptive Mesh Refinement and
a Riemann solver based on the Van~Leer flux-splitting algorithm, which solves
nonequilibrium ionization rate equations (for H, He, C, N, O, Ne, and~S)
and includes an explicit treatment of radiative transfer
which takes account of bound-free opacity of H and He
(\cite{X9},\cite{X10},\cite{X11}).
Our grid size in $(r,z)$ was $128\times512$~cells (fully refined).
Figure~1 shows the structure of the cloud $50\,\rm Myr$ after
it was overtaken by the quasar's I-front as it sweeps past the cloud
in the IGM. Since $\ell_S\ll R_c$ initially, the cloud traps the I-front,
as described above, and drives a supersonic wind from the surface facing the
quasar. It takes more than $100\,\rm Myr$ to evaporate the cloud,
accelerating it to 10's of $\rm km\,s^{-1}$ in the process.
Figure~2 shows selected observable diagnostics, including column densities
of H~I, He~I and II seen along the symmetry axis at different times and the
spatial variation of the relative abundances of selected metal ions at
$50\,\rm Myr$. The cloud starts as a high-column-density Lyman Limit absorber,
but ends with the H~I column density of a Lyman alpha forest cloud, with
$\rm[He\,II]/[H\,I]\sim10^2$ and metal ions.

\begin{figure}
\centerline{\vbox{
\hbox{\hskip0.9cm\psfig{figure=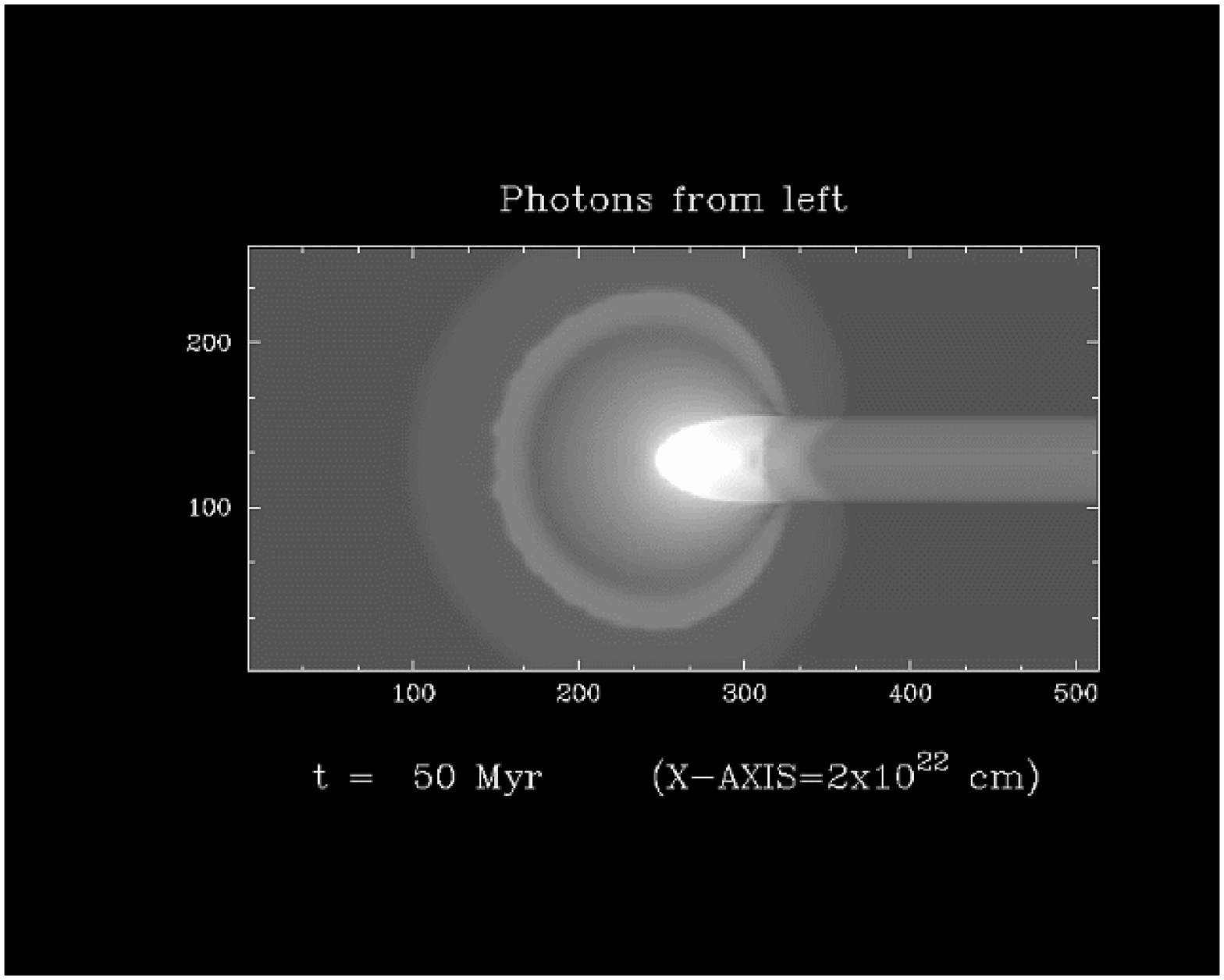,height=7.cm}}
\vskip-1cm
\hbox{\psfig{figure=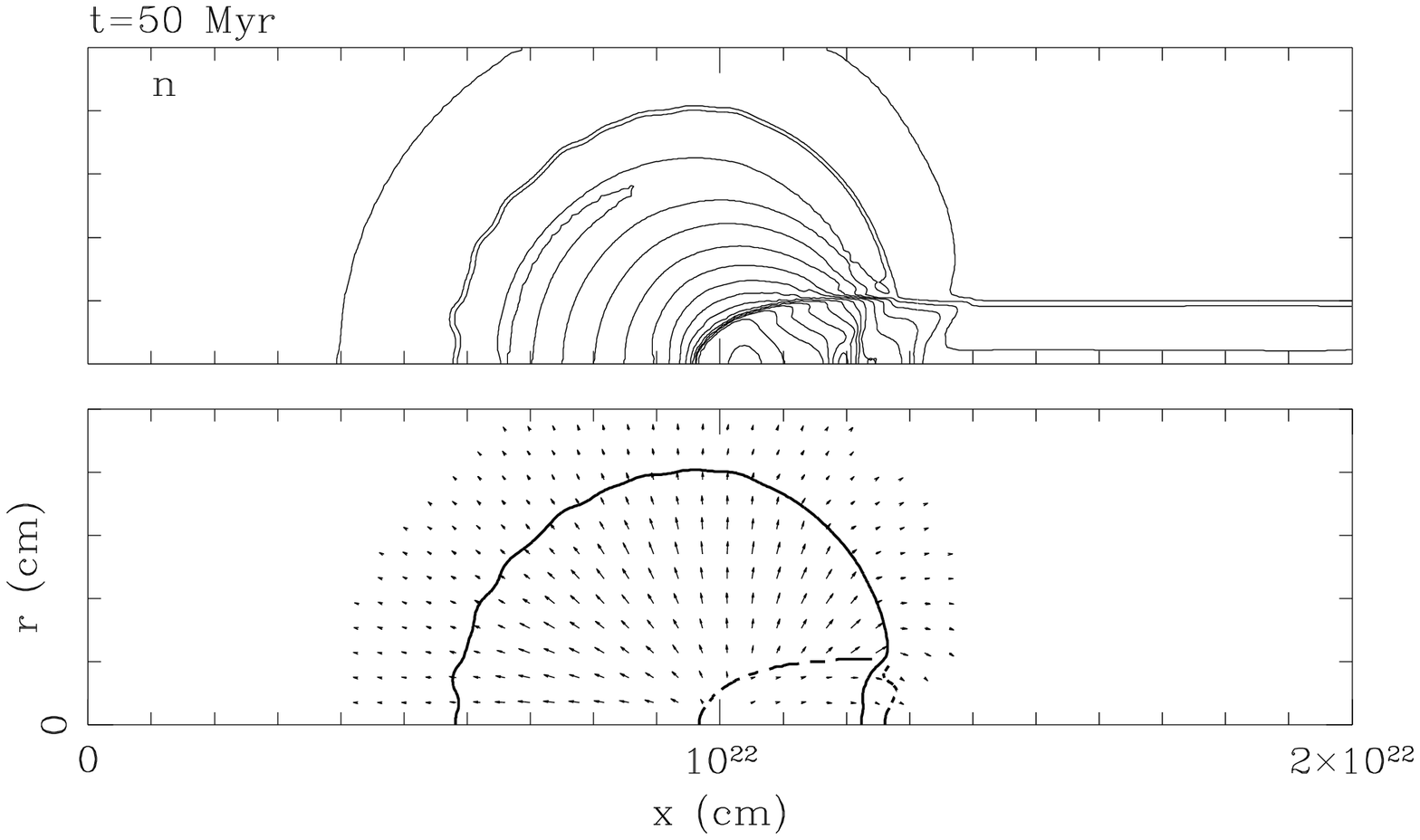,width=10.cm}}
}}
\vskip-3cm
\caption[]
{
THE PHOTOEVAPORATION OF AN INTERGALACTIC GAS CLOUD BY IONIZING
RADIATION FROM A NEARBY QUASAR. (a) Results $50\,\rm Myr$ after
the quasar, located 
$1\,\rm Myr$ away along the x-axis to the
left of the computational box, first turns on.
(a) (upper box) Shaded isodensity
contours with logarithmic spacing, of the total atomic (HI) density $n$
(highest = white, lowest = black).
(b) (lower panels) contour plots of
atomic density (upper), logarithmically spaced, and (lower) velocity
arrows are plotted for velocities larger than 5 km/s, with length 
proportional to velocity. The solid line shows current extent of the original
cloud matter and dashed line is the I-front (50\% H ionization contour).
}
\end{figure}

\begin{figure}
\centerline{\vbox{
\psfig{figure=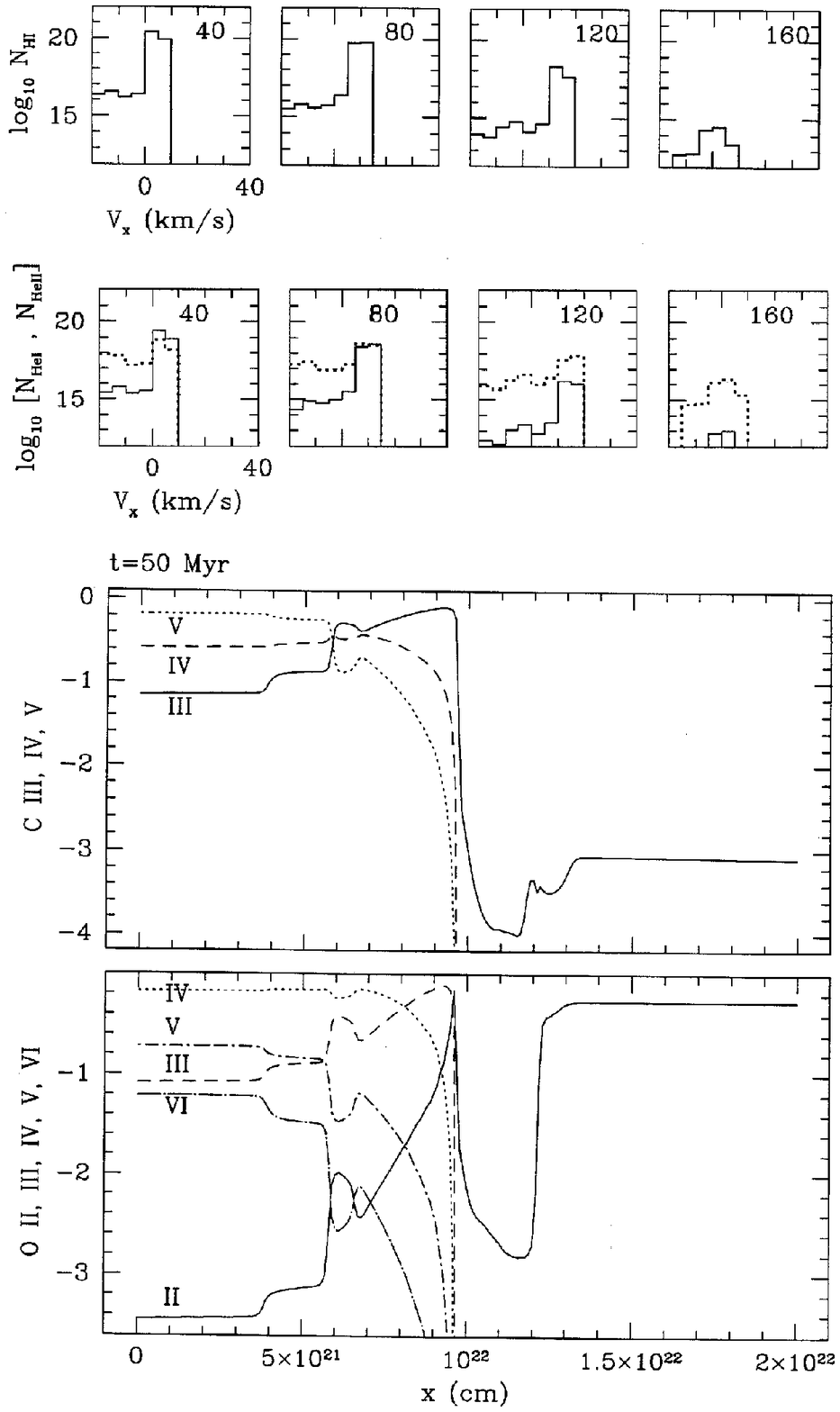,height=16.cm}
}}
\vskip-1cm
\caption[]{PHOTOEVAPORATING CLOUD: OBSERVATIONAL DIAGNOSTICS.
(a) Column Densities of H~I (top panels) and of He I (solid) and II (dotted)
(middle panels) versus velocity as measured along the $x$-axis
at $r=0$. Each box labelled with time (in Myrs) since the QSO turned on.
(b) (bottom panels)
Carbon and Oxygen ionic fractions along this symmetry axis at
$t=50\,\rm Myr$.
}
\end{figure}

\acknowledgements{This work was supported by NASA Grant NAG5-2785 
and NSF Grant ASC-9504046, and was made possible by a UT Dean's
Fellowship and a National Chair of Excellence, UNAM, Mexico in 1997 for PRS.
PRS is also grateful to Hugo Martel and Mark Giroux
for their collaboration in the work referred to in \S1, which
space did not permit us to include here.}

\vfill

\begin{iapbib}{99}{
\bibitem{X}  Shapiro, P. R., Giroux, M. L., \& Babul, A. 1994, ApJ, 427, 25.
\bibitem{X2} Giroux, M. L., \& Shapiro, P. R. 1996, ApJ Suppl., 102, 191.
\bibitem{X3} Shapiro, P. R. 1995, in {\it The Physics of the Interstellar
             Medium}, eds. A. Ferrara, C. F. McKee, C. Heiles, and 
             P. R. Shapiro (ASP Conf. Vol. 80), 55--97.
\bibitem{X4} Shapiro, P. R., \& Martel, H. 1995, in {\it Dark Matter}, eds.
             S. S. Holt and C. L. Bennett (AIP Conf. Proc. 336), pp.~446--449.
\bibitem{X5} Shapiro, P. R., \& Martel, H. 1997, in preparation.
\bibitem{X6} Shapiro, P. R. 1986, PASP, 98, 1014.
\bibitem{X7} Shapiro, P. R., \& Giroux, M. L. 1987, ApJ, 321, L107.
\bibitem{X8} Spitzer, L. 1978, {\it Physical Processes in the Interstellar
             Medium} (Wiley).
\bibitem{X9} Mellema, G., Raga, A. C., Canto, J., Lundqvist, P., 
             Balick, B., Steffen, W., \& Noriega-Crespo, A. 1997, A\&A, 
             submitted.
\bibitem{X10} Raga, A. C., Mellema, G., \& Lundquist, P. 1977, 
              ApJ Suppl., 109, 517.
\bibitem{X11} Raga, A. C., Taylor, S. D., Cabrit, S., \& Biro, S. 1995, 
              A\&A, 296, 833.
}
\end{iapbib}
\end{document}